\documentclass[journal]{IEEEtran}

\usepackage{cite}

\usepackage{wrapfig}
\usepackage{calc}
\usepackage{graphicx}

\begin{document}

\pagenumbering{gobble}
\setcounter{page}{0}
\begin{minipage}[t]{\textwidth}
\fontsize{12pt}{12pt}\selectfont 
\textcopyright{} 2021 IEEE. Personal use of this material is permitted. Permission from IEEE must be obtained for all other uses, in any current or future media, including reprinting/republishing this material for advertising or promotional purposes, creating new collective works, for resale or redistribution to servers or lists, or reuse of any copyrighted component of this work in other works. \\

Version accepted for publication in IEEE Systems, Man, and Cybernetics Magazine.
\end{minipage}

\bstctlcite{IEEEexample:BSTcontrol}

\title{Self-Organizing Software Models for the Internet of Things}

\author{Damian Arellanes% <-this % stops a space
\thanks{D. Arellanes is with the School of Computing and Communications, Lancaster University, Lancaster
LA1 4YW, UK (e-mail: damian.arellanes@lancaster.ac.uk).}% <-this % stops a space
}

\markboth{}
{}

\maketitle

\pagenumbering{arabic}

\begin{abstract}
The Internet of Things (IoT) envisions the integration of physical objects into software systems for automating crucial aspects of our lives, such as healthcare, security, agriculture, and city management. Although the vision is promising, with the rapid advancement of hardware and communication technologies, IoT systems are becoming increasingly dynamic, large, and complex to the extent that manual management becomes infeasible. Thus, it is of paramount importance to provide software engineering foundations for constructing autonomic IoT systems. In this paper, we introduce a novel paradigm referred to as \emph{self-organizing software models} in which IoT software systems are not explicitly programmed, but emerge in a decentralized manner during system operation, with minimal or without human intervention. We particularly present an overview of these models by including their definition, motivation, research challenges, and potential directions.
\end{abstract}


\section{Introduction}

\IEEEPARstart{T}{he} Internet of Things (IoT) is a novel paradigm, considered the next Industrial Revolution, that promises the integration of every physical object for the automation of essential concerns of our modern lives such as healthcare, security, agriculture, and city management. Unlike traditional enterprise systems, IoT systems are moving towards environments full of complex interactions, as a consequence of the overwhelming number of objects available worldwide~\cite{janiesch_internet_2020,arellanes_analysis_2018,fortino_agents_2016}. Currently, there are over 17 billion connected objects, and it is estimated that this number will increase exponentially in the coming years~\cite{arellanes_evaluating_2020}. Hence, scalability and complexity become a significant challenge for the full realization of IoT.

Most IoT research has extensively focused on hardware and network issues so that early IoT systems operate in closed environments and integrate relatively few static software components. Contrastingly, future software-intensive IoT systems will be deployed in open environments (i.e., software ecosystems~\cite{broring_enabling_2017}) where billions of (off-the-shelf) components will abstract the functionality of an immense number of connected physical objects~\cite{arellanes_evaluating_2020,want_enabling_2015}. Such environments will be highly dynamic and uncertain due to disturbances caused by external perturbations (e.g., change in requirements and increasing workloads) and unforeseeable internal situations (e.g., system failures and sub-optimal behaviors) ~\cite{arellanes_evaluating_2020,weyns_software_2019}. 

Autonomicity represents the most viable solution to manage complex IoT systems that both integrate an ultra-large number of software components and operate in highly dynamic, uncertain environments. This is because that property enables the adaptation of computational behaviors with minimal or no human intervention. Research on autonomic software has produced significant results, especially in the area of self-adaptive software~\cite{weyns_software_2019}. However, existing solutions are mainly centralized, and it has been proven over many years that centralized approaches do not scale and are therefore unsuitable to tackle complexity~\cite{heylighen_meaning_2003}. 

In this paper, we present \emph{self-organizing software models} which are a new kind of abstractions that allow the construction of autonomous software systems, in which computational behaviors are not predefined but emerge during system operation to dynamically accommodate a given context. Emergence is achieved from the individual interactions of the constituent software components (not hardware devices), without the need of a central authority. The main role of these models is to remove or reduce the role of a programmer in the composition, maintenance and evolution processes of a software system. 

\section{What is a Self-Organizing Software Model?}

Self-organization is the bottom-up process by which complex behaviors emerge from the decentralized interactions of participant components (e.g., molecules or insects), in order to collectively achieve a global system goal (e.g., foraging). In contrast to top-down processes, self-organization is a well-known technique to deal with uncertainty, scale, dynamism, and complexity~\cite{ashby_principles_1962,heylighen_meaning_2003}. Self-organization is not a new concept. It has been studied in diverse areas from distinct points of view, from biological systems (e.g., flock of birds, school of fish, and ant colonies) to artificial systems (e.g., traffic light ensembles, networking, and swarm robotics). Just recently, self-organization has captured the attention of the software engineering community to study it as an inherent property of software models.

A software model is an abstract system representation that describes software components and their composition~\cite{lau_introduction_2017}. A software component (e.g., a web service or a generic port-based component) is a self-contained unit of composition that provides some computational functionality. In traditional software engineering, composition is the design-time process of combining the functionality of two or more software components, and it is performed manually by system engineers~\cite{lau_introduction_2017,lehman_softwares_1998}. As manual composition is unsuitable to tackle the imminent challenges that future IoT systems pose (i.e., dynamicity, complexity, and scale), we envision that such an approach will eventually be obsolete. Instead, composition will be a run-time process performed by the model itself (i.e., autonomously), in which complex computational behaviors emerge in the form of complex composite components. These emergent composites can further self-organize to define even more complex composites. Emergence occurs from the decentralized interactions of the available autonomous components (potentially developed independently by different stakeholders), with no or minimal 

\begin{wrapfigure}[14]{r}[.5\width+.5\columnsep]{5cm}\itshape
``A self-organizing software model is a computational abstraction whose software components comply with self-organization rules, so complex composite components are not explicitly programmed but (autonomously) emerge from the decentralized interactions of the available, independent software components. Its overall goal is to accommodate perturbations in a system operating environment.''
\end{wrapfigure}

\noindent human intervention and according to a set of self-organization rules. As emergent composites cannot be expressed as a simple summation of the composed components, emergent computational behaviors cannot be predicted just by knowing the available components. 

\begin{figure}[!b]
  \centering
  \includegraphics[scale=0.46]{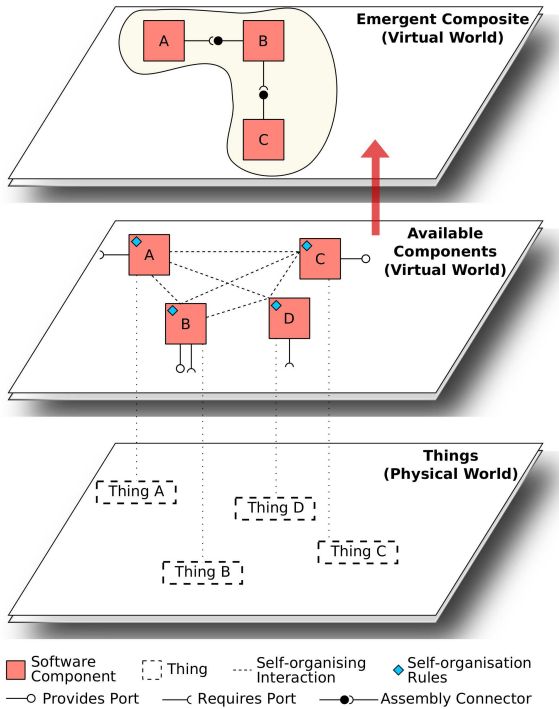} 
  \caption{Self-organizing IoT.}
  \label{fig:self-organising-software-models}
\end{figure}

Our vision of self-organizing software models is depicted in Figure~\ref{fig:self-organising-software-models}. Although we use a port-based composite for illustrative purposes, other software compositions can emerge as a result of self-organizing component interactions, such as service-oriented workflows~\cite{ben_mahfoudh_learning-based_2020,chen_goal-driven_2018} or algebraic compositions~\cite{arellanes_exogenous_2017,arellanes_workflow_2019}. In any case, emergent composites lie on top of a three-layer IoT view, physical objects (known as \emph{things}) are situated at the bottom and self-organizing interactions occur in the middle. This conceptual three-layer view shall be referred to as \emph{self-organizing IoT}.

The idea of self-organizing software contrasts with that of self-adaptive systems. This is because the latter often require adaptation managers to control the entire adaptation process outside a software model~\cite{weyns_software_2019}. We refer to this process as exogenous adaptation. By contrast, in our vision, adaptation is achieved through self-organization, and it therefore occurs in a pure decentralized manner without the need of any central authority, external controllers, or leaders. We refer to this process as endogenous adaptation. Unlike self-adaptive software, in self-organizing software models there is no notion of \emph{managed system} or \emph{managing system} since components, which belong to the system itself, collaboratively realize adaptation by the emergence of computational behaviors. Like self-adaptive 

\begin{wrapfigure}[14]{l}[.5\width+.5\columnsep]{5cm}\end{wrapfigure}

\noindent software, there are adaptation triggers to initiate the emergence of composite components upon detecting perturbations in the internal or the external system operating environment. Perturbations in the internal environment include system failures and sub-optimal performance. Perturbations in the external environment include change in stakeholder requirements, run-time scaling (i.e., component addition/removal), environmental changes, and increasing workloads~\cite{arellanes_evaluating_2020,weyns_software_2019}.

\begin{figure}[h]
  \centering
  \includegraphics[scale=0.45]{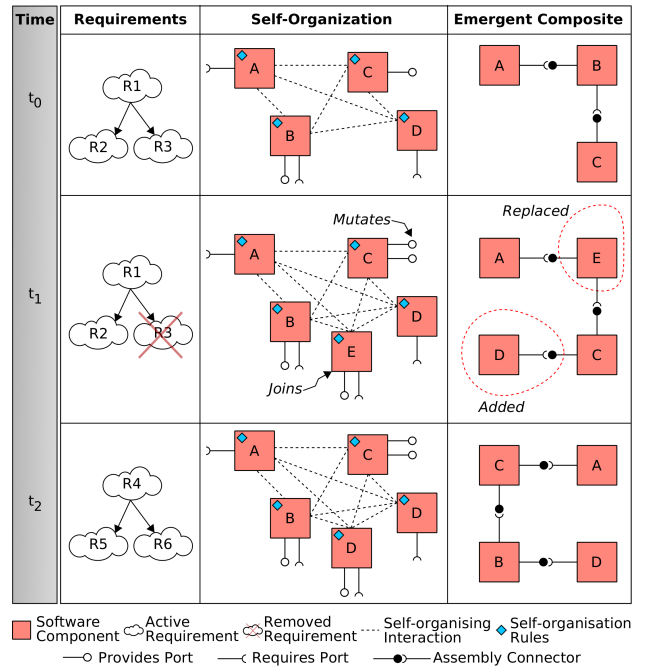} 
  \caption{An example of a self-organizing software model.}
  \label{fig:example}
\end{figure}

To illustrate how self-organizing software components deal with open environments, let us consider the example depicted in Figure~\ref{fig:example}. At time $t_0$, four components interact to meet the requirements \emph{R1}, \emph{R2}, and \emph{R3}. In this case, the result of self-organization is a composite assembling \emph{A}, \emph{B}, and \emph{C}. Suddenly, at time $t_1$, the requirement \emph{R3} is no longer needed, and component \emph{E} becomes available. As requirements changed and component \emph{E} offers better performance than \emph{B}, the existing components self-organize to connect \emph{C} with \emph{D} and replace \emph{B} with \emph{E}. Note that component \emph{C} has mutated by adding a new port, in order to fulfill the requirements \emph{R1} and \emph{R2}. Finally, in the last time window, the requirements \emph{R1}, \emph{R2}, and \emph{R3} become obsolete since a new specification is defined. So, components self-organize once again to compose an entirely new computational structure. In any time window, composite components emerge without any central controller or leader.

\section{Why Self-Organizing Software Models?}
\label{sec:motivation}

With the rapid advancement of hardware and communication technologies, IoT systems are becoming increasingly dynamic, functionally large, and extremely complex to the extent that manual management becomes infeasible. Dynamism, 

\begin{wrapfigure}[7]{r}[.5\width+0.5\columnsep]{5cm}\itshape
``Self-organizing software models allow the evolution of large-scale IoT software systems which operate in environments with a high degree of uncertainty, complexity, and dynamism.''
\end{wrapfigure}

\noindent or churn~\cite{arellanes_evaluating_2020}, makes \emph{things} (and their software components) to constantly appear and disappear in completely uncertain network environments~\cite{chen_goal-driven_2018}. This can be a consequence of mobility, failures, or poor network connections. Functional scalability~\cite{arellanes_evaluating_2020} is another problem, which arises from the fact that the functionality of one \emph{thing} can be virtualized by more than one software component, and there are plenty of \emph{things} available. In fact, the more components available, the more possible computational behaviors (potentially leading to a combinatorial explosion problem~\cite{arellanes_exogenous_2017,arellanes_workflow_2019}). Last but not least, complexity refers to the number of interactions between software components~\cite{arellanes_analysis_2018,heylighen_meaning_2003,lehman_softwares_1998}, and it is closely related to functional scalability since the more software components composed, the more complex a system is. To tackle these imminent challenges, it is therefore of paramount importance to provide software engineering theory that facilitates the construction of autonomic IoT software systems. 

Most of the research done in the field of autonomic computing is built upon centralized adaptation managers (e.g., MAPE-K control loop~\cite{kephart_vision_2003}). However, due to the law of requisite variety~\cite{ashby_requisite_1958}, which refers to an explosion in the number of system states, such top-down autonomic solutions are unsuitable for dealing with realistic, open environments like IoT~\cite{gershenson_guiding_2020,georgiadis_self-organising_2002}. 

Since self-organization is a well-known bottom-up approach that deals with precisely the challenges that IoT faces~\cite{ashby_principles_1962}, some preliminary endeavors have been done to apply self-organization principles in the IoT realm. However, most of this work has been done from the perspective of general systems engineering (e.g.,~\cite{gershenson_guiding_2020}) rather than from a software engineering viewpoint. Just a few works have been devised in the context of self-organizing software models~\cite{ben_mahfoudh_learning-based_2020,chen_goal-driven_2018,bures_software_2016,cardozo_emergent_2016,dowling_self-managed_2004,georgiadis_self-organising_2002}. 

Because IoT systems are becoming increasingly software-intensive, we need to define self-organization rules in the semantics of software components. Otherwise, self-organization does not occur among software components but among something else (e.g., network nodes~\cite{mamei_emergence_2005} or robots~\cite{baldassarre_self-organized_2007}). We need to leverage the flexibility and manageability that software offers versus hardware (cf., software-defined networks~\cite{hayek_analysis_2018}) in order to adapt IoT systems to different operation contexts. In the end, it is software that controls embedded systems in \emph{things}, and it is because of this that self-organization always originates in the virtual world. Embedded systems are just interfaces to the physical reality.

As they provide the mechanisms to compose/emerge complex computational behaviors on the fly without any central adaptation entity, we see self-organizing software models as an important contribution to the field of autonomic computing. 

\section{Research Challenges and Potential Directions}
\label{sec:challenges}

Due to its non-deterministic nature, the main challenge of self-organization is to emerge meaningful functionality from decentralized interactions. As the concept of meaningful varies 

\begin{wrapfigure}[8]{l}[.5\width+.5\columnsep]{5cm}\end{wrapfigure}

\noindent from one domain to another, goal models (e.g., RELAX~\cite{whittle_relax_2010} and KAOS~\cite{van_lamsweerde_managing_1998}) can be used to specify expected system-wide behaviors in the form of \emph{requirements@run-time} (which must be met without any central reasoner). In addition to goal models, run-time verification/testing techniques can be integrated to prove the correctness of emergent computational behaviors. All of this without sacrificing systems' operation. Apart from dealing with \textit{non-determinism} and \textit{correctness}, other challenges are as follows:

\begin{itemize}
\item \textit{Incompleteness}. Due to their decentralized nature, software components cannot always have a complete, consistent view of their operating environment (e.g., knowledge about other available components or knowledge about running composition structures). This is especially true in IoT ecosystems with a large number of IoT software components. So, How to efficiently disseminate knowledge to maintain an accurate representation of the world under highly dynamic environments?
\item \textit{Self-explanation}. Transparency of decision-making is a must in approaches lacking human intervention, since users may require an explanation of the emergence of certain computational behavior. So, How to provide self-explanation in self-organizing software models, especially when the participant components are completely autonomous and operate in different administrative domains?
\item \textit{Measurement}. To date, there is a huge body of research on self-organizing systems in which plenty of different evaluation metrics have been proposed, e.g., entropy, fragility, and stability~\cite{gershenson_guiding_2020}. Are these metrics suitable to evaluate computational emergence in self-organizing software models? If not, what are the most suitable metrics for the software engineering domain?
\item \textit{Uncertainty quantification}. Since IoT systems operate in environments full of \emph{aleatoric} and \emph{epistemic} uncertainty, it is impossible to predict the space of complex programs that arise from dynamically appearing and disappearing software components. To control the construction of such spaces, uncertainty quantification approaches can be used for defining trustworthy decision-making mechanisms that proactively adjust the way components interact. What are the semantic constructs that allow software components to collaboratively solve uncertainty quantification problems? 
\item \textit{Dynamic evolution of self-organization rules}. Self-organization rules can be predefined at design-time. However, in certain scenarios, the rules might not be enough to achieve global system goals. So, Is it possible to define decentralized learning techniques (e.g., collaborative/multi-agent reinforcement learning) for dynamically evolving self-organization rules without human intervention? 
\item \textit{Evolution reasoning}. Self-organizing software models are always evolving at run-time, so it becomes necessary to reason about dynamics. Modeling evolution in traditional self-organizing systems can be done using process algebra, temporal logic, or Petri Nets. But are these techniques suitable to reason about self-evolving software? If not, which reasoning techniques could be more appropriate?
\item \textit{Software semantics and self-organization rules}. Defining generic self-organization rules for autonomously composing software in different IoT domains is challenging. This raises the question of what are the most suitable self-organization rules for IoT software models? Can we embed those rules in the semantics of a (universal) component model? 
Can we take inspiration from self-organizing biological systems to define bio-inspired component models? 
\end{itemize}

\section{Conclusion}

Although the dream of removing the tasks performed by software engineers is far from reality, self-organization offers an encouraging route towards the next generation of software systems in which software is not explicitly programmed, but emerges without any central controller to meet the needs of a given context at run-time. We refer to this class of abstractions as \emph{self-organizing software models}. In this paper, we presented the definition, motivation, challenges and future directions of these abstractions. 

To date, self-organizing software models are still in their infancy and their challenges hinder the development of a self-organizing software solution applicable to a wide variety of IoT domains. We envision that this nascent field will be of great relevance in the coming years to deal with the inherent scale, uncertainty, dynamism, and complexity that IoT systems are increasingly posing.

\bibliographystyle{IEEEtran}
\bibliography{IEEEabrv,refs}

\begin{IEEEbiographynophoto}{Damian Arellanes} (damian.arellanes@lancaster.ac.uk) is an Assistant Professor in Computer Science at the School of Computing and Communications, Lancaster University, UK. He received his PhD degree in Computer Science from the University of Manchester, UK. His research interests lie at the intersection of models of computation, Cyber-Physical Systems, and autonomic computing, with a particular emphasis on self-organizing software models.
\end{IEEEbiographynophoto}

\end{document}